\documentclass[a4paper]{jpconf}
\usepackage{graphicx}
\begin{document}
\title{Magnetization study on the field-induced quantum critical point in YbRh$_2$Si$_2$}

\author{Y. Tokiwa$^{1,2}$, C. Geibel$^2$, F. Steglich$^2$, and P. Gegenwart$^{1}$}

\address{
$^1$I. Physik. Institut, Georg-August-Universit\"{a}t G\"{o}ttingen, D-37077 G\"{o}ttingen\\
$^2$Max-Planck Institute for Chemical Physics of Solids, D-01187 Dresden, Germany}

\ead{ytokiwa@gwdg.de}

\begin{abstract}
We study the field-induced quantum critical point (QCP) in
YbRh$_2$Si$_2$ by low-temperature magnetization, $M(T)$, and
magnetic Gr\"uneisen ratio, $\Gamma_{\rm mag}$, measurements and
compare the results with previous thermal expansion, $\beta(T)$, and
critical Gr\"uneisen ratio, $\Gamma^{cr}(T)$, data on
YbRh$_2$(Si$_{0.95}$Ge$_{0.05}$)$_2$. In the latter case, a slightly
negative chemical pressure has been used to tune the system towards
its zero-field QCP. The magnetization derivative $-dM/dT$ is far
more singular than thermal expansion, reflecting a strongly
temperature dependent pressure derivative of the field at constant
entropy, $(dH/dP)_S=V_m\beta/(dM/dT)$ ($V_m$: molar volume), which
saturates at $(0.15\pm 0.04)$~T/GPa for $T\rightarrow 0$. The line
$T^\star(H)$, previously observed in Hall- and thermodynamic
measurements, separates regimes in $T$-$H$ phase space of stronger
$(\epsilon>1$) and weaker $(\epsilon<1$) divergent $\Gamma_{\rm
mag}(T)\propto T^{-\epsilon}$.
\end{abstract}

\section{Introduction}
Quantum criticality in YbRh$_2$Si$_2$ (YRS) has recently attracted
considerable interest because of divergences of the quasiparticle
(qp) effective mass and qp-qp scattering
cross-section~\cite{GegenwartP:Magiqc,CustersJ:brehea}, as well as a
drastic change of the Hall-coefficient \cite{PaschenS:Haleah} upon
tuning through the quantum critical point (QCP), which is also
accompanied by signatures in thermodynamic and magnetic properties
\cite{GegenwartP:Mulesa}. Undoped YRS at ambient pressure shows very
weak antiferromagnetic (AF) order below $T_{\rm N}=70$~mK, which
could be tuned continuously down to $T=0$ by either application of a
small magnetic field or by a tiny volume expansion (negative
chemical pressure) in YbRh$_2$(Si$_{0.95}$Ge$_{0.05}$)$_2$.
Important advantages of using magnetic field compared to chemical
pressure for tuning the system towards the QCP are that fields could
be changed continuously and no disorder is introduced by doping.\\
Recently, the Gr\"uneisen ratio, $\Gamma\propto \beta/C$, where
$\beta$ and $C$ denote the volume thermal expansion and specific
heat, respectively, has been identified as the most sensible probe
of the nature of quantum criticality for {\it pressure}-driven QCPs
and correspondingly, the magnetic Gr\"uneisen ratio, $\Gamma_{\rm
mag}=-(dM/dT)/C$ ($M$: magnetization) for the {\it field}-induced
case \cite{ZhuLijun:UnidGp,GarstM:SigctG}. For Ge-doped YRS, the
critical contribution to the Gr\"uneisen ratio has been found to
diverge with fractional exponent, $\Gamma^{cr}\propto T^{-0.7}$,
below 0.3~K~\cite{KuchlerR:DivtGr}. This temperature dependence is
in strong disagreement with the prediction of the traditional
scenario for an itinerant AF QCP~\cite{ZhuLijun:UnidGp} and has been
taken as evidence for local quantum criticality in this
system~\cite{KuchlerR:DivtGr}. Very recently, a detailed study of
the magnetic Gr\"uneisen ratio in the $T$-$H$ plane of the phase
diagram of undoped YRS has been reported~\cite{ISI:000263389500045}.
$\Gamma_{\rm mag}$ diverges when approaching the QCP either at
$H=H_c$ as a function of temperature or within the Landau-Fermi
liquid regime (as $T\rightarrow 0$) at $H>H_c$, when tuning the
field towards $H_c$. For the latter case, $\Gamma_{\rm
mag}(T\rightarrow 0)=-0.3/(H-H_c)$ has been found. Such a
($H-H_c)^{-1}$ dependence is expected from a scaling ansatz for a
field-driven QCP. Furthermore, the prefactor should equal the
exponent of the divergence of the specific heat coefficient, which
is indeed fulfilled \cite{ISI:000263389500045}. In contrast to this
rather clear situation in the approach of the QCP at $H>H_c$ by
reduction of the field towards $H_c$, more complicated behavior has
been found for the temperature dependence at $H$=$H_c$ \cite{
ISI:000263389500045}. At high temperatures, an unexpectedly strong
divergence $\Gamma_{\rm mag}\propto T^{-2}$ is found with a
crossover at about 0.3~K towards a weaker $T^{-0.7}$ divergence. In
this paper, we discuss how this complicated behavior may arise as
consequence of the additional low-energy scale $T^\star(H)$,
previously observed in Hall- and thermodynamic
measurements~\cite{PaschenS:Haleah,GegenwartP:Mulesa}. We also
investigate the relation between chemical-pressure and field tuning
in YRS by comparing the temperature derivative of the magnetization
with the volume thermal expansion.

\section{Experimental}
High-quality single crystals ($\rho_0=1~\mu\Omega$cm) were grown
from In-flux as described earlier~\cite{trovarelli:prl-00}. The DC
magnetization was measured utilizing a high-resolution capacitive
Faraday magnetometer~\cite{SAKAKIBARAT:Farfmh}. In this paper, we
analyze magnetic Gr\"uneisen parameter results from
\cite{ISI:000263389500045} and compare them with thermal expansion
data from~\cite{KuchlerR:DivtGr}.

\section{Results and discussion}

\begin{figure}[h]
\begin{minipage}{18pc}
\includegraphics[width=18pc]{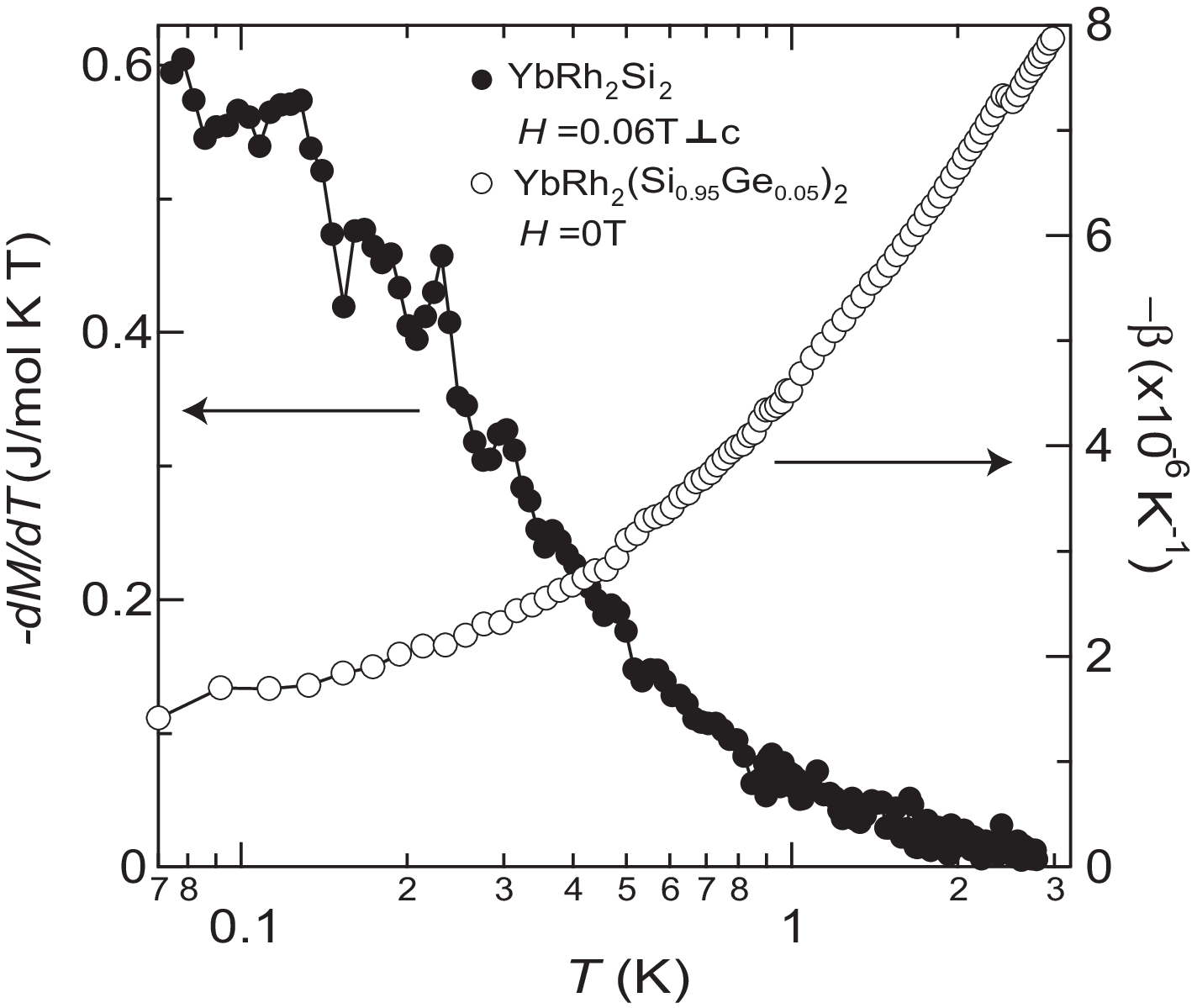}\hspace{2pc}%
\caption{\label{dMdTbeta}Temperature derivative of the magnetization
$-dM/dT$ of YbRh$_2$Si$_2$ at $H$=0.06\,T (left axis) and volume
thermal expansion $-\beta$ of
YbRh$_2$(Si$_{0.95}$Ge$_{0.05}$)$_2$~\cite{KuchlerR:DivtGr} at zero
field (right axis) as a function of temperature (on a logarithmic
scale).}
\end{minipage}\hspace{2pc}%
\begin{minipage}{16pc}
\includegraphics[width=14pc]{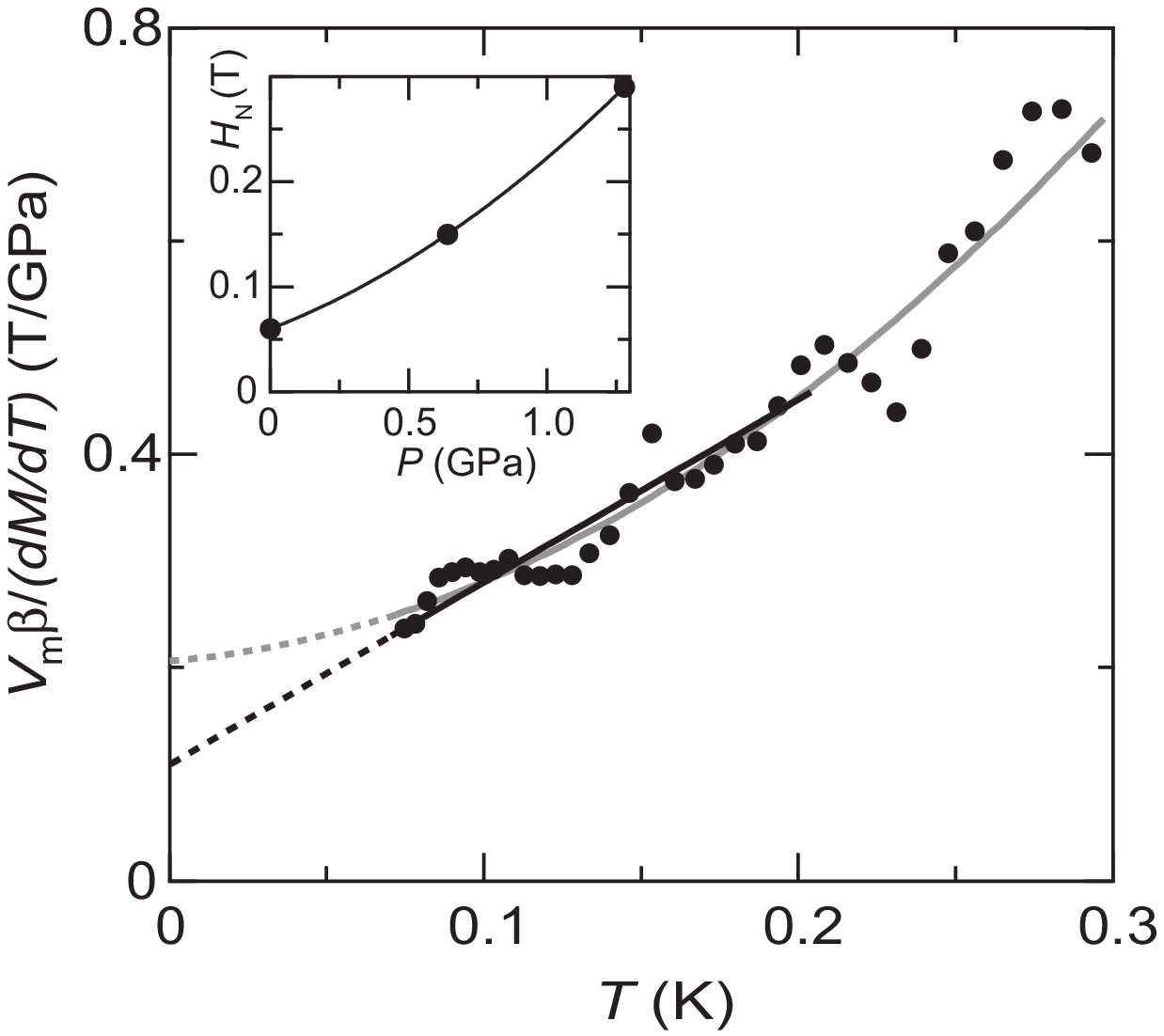}\hspace{2pc}%
\caption{\label{ratio}$V_m\beta/(dM/dT)$ as a function of
temperature. Grey and black lines indicate a third polynomial and a
linear dependence (fitted below 0.3 and 0.2\,K), respectively.
Inset: Pressure dependence of AF critical field of
YbRh$_2$Si$_2$~\cite{TokiwaY:Fiesth}. Solid line represents
$H_N(P)=a+bP+cP^2$ with initial slope $b=0.11$~T/GPa.}
\end{minipage}
\end{figure}

At first, we compare the temperature derivative of the
magnetization, which equals the field derivative of the entropy, for
undoped YRS at $H=H_c$ with the volume thermal expansion for
Ge-doped YRS near the critical chemical pressure. The latter
property equals the pressure-derivative of the entropy, so the ratio
$V_m\beta/(dM/dT)=(dH/dP)_S$ ($V_m$: molar volume) equals the
pressure derivative of the field at constant entropy. A comparison
of the two former properties is shown in Figure~1, whereas their
ratio is displayed in the main part of Figure~2. This ratio is
temperature dependent and decreases with decreasing $T$,
extrapolating to a value of $(0.15\pm 0.04)$~T/GPa where the lower
and upper boundaries are given by either a linear or a third
polynominal extrapolation as indicated by the black and gray lines,
respectively. This value could be compared with a measurement of the
pressure dependence of the critical field of AF ordering in YRS,
derived from magnetization measurements under hydrostatic
pressure~\cite{TokiwaY:Fiesth}. As shown in the inset of Figure~2,
the initial pressure derivative of the N\'{e}el field is derived as
0.11~T/GPa which is close the lower bound of $(dH/dP)_S$ derived
from the ratio of thermal expansion to $dM/dT$, proving the
consistency of our thermodynamic data.

The temperature dependences of $dM/dT$ for undoped YRS at $H=H_c$
and $\beta(T)$ at $H=0$ for Ge-doped YRS are quite different, in
particular at high temperatures. Whereas the former strongly
increases upon cooling to below 4~K, reverse behavior is found in
the latter property, resulting in a strongly temperature dependent
$(dH/dP)_S$. For the analysis of the critical Gr\"uneisen ratio
$\Gamma^{cr}$ at low temperatures \cite{KuchlerR:DivtGr} a
substantial constant "background contribution" in thermal expansion
divided by temperature has been subtracted whereas for the study of
the {\it magnetic} Gr\"uneisen ratio no background in $(dM/dT)/T$
needs to be subtracted, since a possible constant contribution is
negligibly small~\cite{ISI:000263389500045}. Both, the critical
"thermal" Gr\"uneisen ratio $\Gamma^{cr}(T)$, as well as the
magnetic Gr\"uneisen ratio, $\Gamma_{\rm mag}(T)$, display a
crossover towards a $T^{-0.7}$ divergence below 0.3~K, which for the
former property has been taken as evidence for local quantum
criticality~\cite{KuchlerR:DivtGr}. The low-$T$ magnetic
susceptibility and specific heat coefficient of YRS also display
crossovers at 0.3~K~\cite{CustersJ:brehea}. The analysis of the
exponent of the magnetic Gr\"uneisen ratio presented in the
following, suggests that these crossovers are related to the
additional low-energy scale $T^\star(H)$.

%Figure 1 displays contrasting behaviors of $-dM/dT$ for YbRh$_2$Si$_2$ at $H$=0.06\,T and $-\beta$ of YbRh$_2$(Si$_{0.95}$Ge$_{0.05}$)$_5$ at zero field. $-dM/dT$ shows a steep increase upon decreasing temperature, while $-\beta$ decreases monotonically. This leads to a much stronger divergence of $\Gamma_{\rm mag}$ than $\Gamma$~\cite{ISI:000263389500045}. The slopes of the both two curves become smaller in low temperature region, resulting in a tendency of saturation in the ratio of the two quantities.

\begin{figure}[h]
\includegraphics[width=30pc]{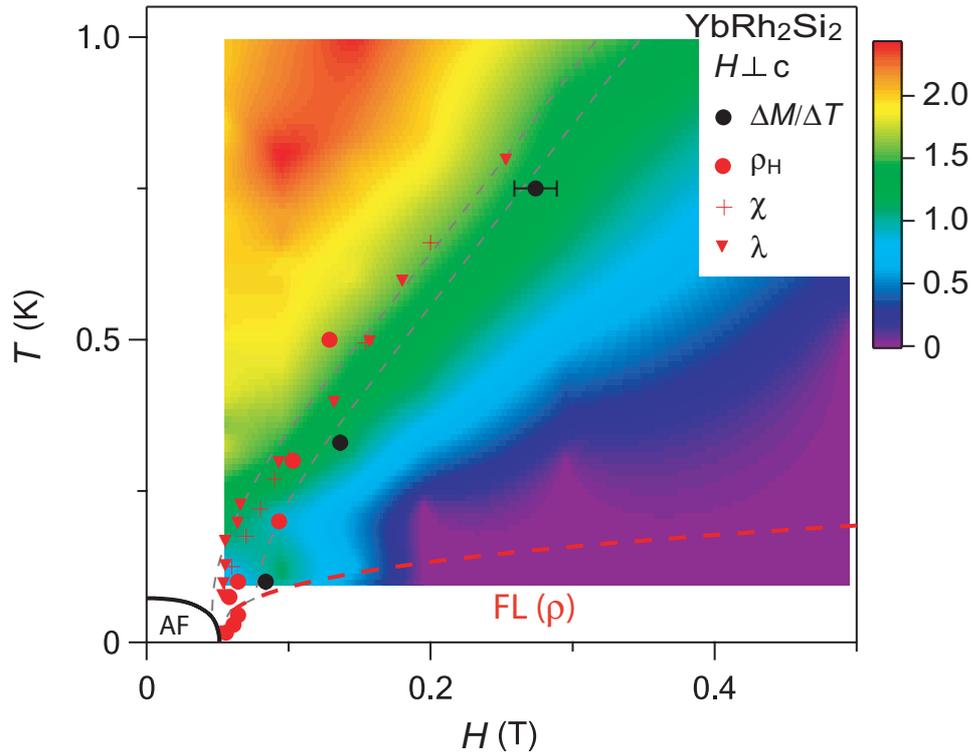}\hspace{2pc}%
\caption{\label{contour}Contour plot in $T$-$H$ phase space for the
power-law exponent of the magnetic Gr\"{u}neisen ratio in
YbRh$_2$Si$_2$ ($H\perp c$). Positions of $T^{\star}(H)$ determined
by $\Delta M/\Delta T(H)$~\cite{ISI:000263389500045}, susceptibility
and magnetostriction~\cite{GegenwartP:Mulesa}, as well as Hall
effect~\cite{PaschenS:Haleah} are plotted for comparison. The black
and red lines represent the boundary of the AF state and the
cross-over towards Fermi-liquid behavior in the electrical
resistivity, respectively~\cite{GegenwartP:Mulesa}.}
\end{figure}

%The ratio between $\beta$ and $dM/dT$ gives the pressure dependence of the QC field as expressed by $dH_c/dp$=$V_m\beta/(dM/dT)$, where $V_m$ is the molar volume~\cite{GarstM:SigctG}. The ratio $V_m\beta/(dM/dT)$ obtained from the data in Fig.~1 (see Fig.~2) is temperature dependent and decreases with decreasing temperature. The estimation of $V_m\beta/(dM/dT)$ for $T\rightarrow$0 depends slightly on extrapolation. Assuming that a linear and a third polynomial fitting give the lower and upper bounds for zero temperature value, $V_m\beta/(dM/dT)(T\rightarrow$0) is estimated to be between 0.11 and 0.19\,T/GPa. The AFM critical field directly measured by magnetization under pressure is plotted in the inset of Fig.~2~\cite{TokiwaY:Fiesth}. A third polynomial fitting yields $dH_{\rm N}/dp$=0.11\,T/GPa at $P$=0\,GPa, which is within the two bounds.

Figure~3 displays the evolution of the magnetic Gr\"uneisen ratio
exponent $\epsilon$, derived from $\Gamma_{\rm mag}(T)\propto
T^{-\epsilon}$, in the temperature-field phase diagram of YRS.
Within the Fermi-liquid regime, $\epsilon=0$ is
predicted~\cite{ZhuLijun:UnidGp,GarstM:SigctG}. We note, that this
regime (cf. the violet colored region in Fig.~3) increases roughly
linearly with increasing $H$ and extends to temperatures about twice
as large than $T_{\rm FL}(H)$ as derived from the resistivity
measurements~\cite{GegenwartP:Mulesa}. The boundary of the Fermi
liquid regime is known to be different for different physical
quantities. \\The scaling analysis predicts a divergence of the
magnetic Gr\"{u}neisen ratio according to $\Gamma_{\rm mag}\propto
{\it T}^{-1/\nu z}$ within the quantum critical regime ($\nu$ and
$z$ are correlation-length exponent and dynamical exponent,
respectively). However, $\Gamma_{\rm mag}(T)$ at $H=0.06$\,T does
not show a single power-law exponent, but rather displays a strong
temperature dependence of the exponent. Most interestingly,
$\epsilon\approx 1$ is found close to $T^\star(H)$, whereas larger
and smaller values are found at $T>T^\star(H)$ and $T<T^\star(H)$,
respectively. The broad region (cf. the green colored region in
Fig.~3) where this change of the exponent takes place suggests that
the 0.3~K crossovers in the various physical properties at $H=0$ and
$H=H_c$~\cite{CustersJ:brehea} may also result as consequence of
$T^\star(H)$. Recently, a critical Fermi surface model for a local
QCP has been proposed, in which the electronic criticality is
described by $\nu=2/3$, $z=3/2$, and $d$=1~\cite{senthil:035103}. In
our previous study, from the measured $\Gamma_{\rm mag}$ we derived
the quantity $G_r=\nu(d-z)=-0.3$, which is in good agreement with
$G_r=-1/3$ from the critical Fermi surface model. Furthermore, this
model predicts $\epsilon$=$1/\nu z=1$. Remarkably, the
$T^{\star}(H)$ line traces a constant-exponent region with
$\epsilon\sim$1.

In conclusion, our comparison of the temperature derivative of the
magnetization for undoped YRS with the thermal expansion of Ge-doped
YRS provides information on the pressure derivative of the field at
constant entropy, $(dH/dP)_S$. This property is strongly temperature
dependent and saturates for $T\rightarrow 0$ at $(0.15\pm
0.04)$~T/GPa. The contour plot of the magnetic Gr\"uneisen ratio
suggests that the temperature dependence within the non-Fermi liquid
region of the $T$-$H$ phase diagram is strongly affected by
$T^\star(H)$, which may explain the crossover scales near 0.3~K
observed in various magnetic and thermodynamic properties. Our
thermodynamic data are compatible with the recently proposed
critical Fermi surface model~\cite{senthil:035103}.

We acknowledge discussions with M. Brando, M. Garst, N. Oeschler, Q.
Si, T. Senthil, and M. Vojta. This work was supported by the DFG
research unit 960 "Quantum phase transitions".

\section{References}
\bibliographystyle{iopart-num}
\bibliography{hf_mod}

\end{document}